\documentclass[]{pasj01}
\usepackage{lineno}
\usepackage{url}
\usepackage{color}

\Received{}%{yyyy/mm/dd}
\Accepted{}%{yyyy/mm/dd}
\begin{document} 

\title{ 
%\LETTERLABEL %%% <-- uncomment for LETTER article  
%\REVIEWLABEL %%% <-- uncomment for REVIEW article  
A Close-in Planet Orbiting Giant Star HD 167768}

%%% begin:list of authors
% Do NOT capitalize all letters in "textsc".
\author{
Huan-Yu \textsc{Teng}\altaffilmark{1,*}, 
Bun'ei \textsc{Sato}\altaffilmark{1}, 
Masanobu \textsc{Kunitomo}\altaffilmark{2}, 
Takuya \textsc{Takarada}\altaffilmark{3}, 
Masashi \textsc{Omiya}\altaffilmark{3,4}, 
Hiroki \textsc{Harakawa}\altaffilmark{5}, 
Guang-Yao \textsc{Xiao}\altaffilmark{6,7} 
Yu-Juan \textsc{Liu}\altaffilmark{6} 
Hideyuki \textsc{Izumiura}\altaffilmark{8}, 
Eiji \textsc{Kambe}\altaffilmark{5}, 
Michitoshi \textsc{Yoshida}\altaffilmark{5}, 
Yoichi \textsc{Itoh}\altaffilmark{9}, 
Hiroyasu \textsc{Ando}\altaffilmark{4}, 
Eiichiro \textsc{Kokubo}\altaffilmark{10}, and 
Shigeru \textsc{Ida}\altaffilmark{11} 
}

\email{teng.h.aa@m.titech.ac.jp}
\altaffiltext{1}{Department of Earth and Planetary Sciences, School of Science, Tokyo Institute of Technology, 2-12-1 Ookayama, Meguro-ku, Tokyo 152-8551, Japan}
\altaffiltext{2}{Department of Physics, Kurume University, 67 Asahi-machi, Kurume, Fukuoka 830-0011, Japan}
\altaffiltext{3}{Astrobiology Center, National Institutes of Natural Sciences, 2-21-1 Osawa, Mitaka, Tokyo 181-8588, Japan}
\altaffiltext{4}{National Astronomical Observatory of Japan, National Institutes of Natural Sciences, 2-21-1 Osawa, Mitaka, Tokyo 181-8588, Japan}
\altaffiltext{5}{Subaru Telescope, National Astronomical Observatory of Japan, National Institutes of Natural Sciences, 650 North A’ohoku Pl., Hilo, HI, 96720, USA}
\altaffiltext{6}{CAS Key Laboratory of Optical Astronomy, National Astronomical Observatories, Chinese Academy of Sciences, Beijing 100101, China} % Liu and Xiao
\altaffiltext{7}{Department of Physics, College of Science, Tibet University, Lhasa 850000, China}
\altaffiltext{8}{Okayama Branch Office, Subaru Telescope, National Astronomical Observatory of Japan, National Institutes of Natural Sciences, Kamogata, Asakuchi, Okayama 719-0232, Japan}
\altaffiltext{9}{Nishi-Harima Astronomical Observatory, Center for Astronomy, University of Hyogo, 407-2, Nishigaichi, Sayo, Hyogo 679-5313, Japan}
\altaffiltext{10}{The Graduate University for Advanced Studies (SOKENDAI), 2-21-1 Osawa, Mitaka, Tokyo 181-8588, Japan}
\altaffiltext{11}{Earth-Life Science Institute, Tokyo Institute of Technology, 2-12-1 Ookayama, Meguro-ku, Tokyo 152-8551, Japan}

%% `\KeyWords{}' always has to be placed before ``\maketitle'' 
%%  List of Key Words:  https://academic.oup.com/pasj/pages/Pasj_Keywords 

\maketitle
\begin{abstract}
We report the detection of a giant planet orbiting a G-type giant star HD 167768 from radial velocity measurements using HIgh Dispersion Echelle Spectrograph (HIDES) at Okayama Astrophysical Observatory (OAO). 
HD 167768 has a mass of $1.08_{-0.12}^{+0.14} M_{\odot}$, a radius of $9.70_{-0.25}^{+0.25} R_{\odot}$, a metallicity of $\rm{[Fe/H]}=-0.67_{-0.08}^{+0.09}$, and a surface gravity of $\log g = 2.50_{-0.06}^{+0.06}$.
The planet orbiting the star is a warm Jupiter, having a period of $20.6532_{-0.0032}^{+0.0032}\ \rm{d}$, a minimum mass of $0.85_{-0.11}^{+0.12}\ M_{\rm{J}}$, and an orbital semimajor axis of $0.1512_{-0.0063}^{+0.0058}\ \rm{au}$. 
The planet has one of the shortest orbital periods among those ever found around deeply evolved stars ($\log g < 3.5$) using radial velocity methods.
The equilibrium temperature of the planet is $1874\ \rm{K}$, as high as a hot Jupiter.
The radial velocities show two additional regular variations at $41\ \rm{d}$ and $95\ \rm{d}$, suggesting the possibility of outer companions in the system. Follow-up monitoring will enable validation of the periodicity.
We also calculated the orbital evolution of HD 167768 b and found that the planet will be engulfed within 0.15\,Gyr.

\end{abstract}

% -------------------------------------------------------------------------------------
\section{Introduction}
Planets around evolved stars have been widely surveyed over the last 20 years, and over 150 planets have been discovered around evolved stars ($\log g < 3.5$)\footnote{NASA Exoplanet Archive \url{https://exoplanetarchive.ipac.caltech.edu} \citep{Akeson2013}}. 
However, the lack of close-in planets around giant stars is well-known in the planet population study.
Theoretically, radial velocity (RV) measurements should be more capable of discovering these close-in planets rather than wide-orbit ones.
However, current surveys reveal that close-in ($a \lesssim 0.6\ \rm{au}$) planets are seldom found around evolved stars (e.g. \cite{Johnson2007, Lillo-Box2016, Medina2018, Teng2022}), while most of the planets survive at distant places to their host stars.

Planet population synthesis suggests the lack of close-in planets could be caused by a scaling of the proto-planetary disk mass with the mass of the central star \citep{Alibert2011}, but the scaling would be the case only if associated with a decrease in the mean disk lifetime for stars more massive than $1.5 M_{\odot}$.
Theoretical studies in stellar evolution suggest the lack of close-in planets could be also attributed to the expansion of central stars ending with planet engulfment (e.g., \cite{Villaver2009, Kunitomo2011, Villaver2014}). 
This scenario would be at least the case for low-mass giant stars ($M \lesssim 1.5 M_{\odot}$), where a large number of short-period planets have been detected around their progenitors, that is, main-sequence stars. 
Furthermore, hot Jupiters may be also destroyed by tidal interactions during the main sequence lifetimes of their host stars \citep{Hamer2019}. 

To date, several close-in planets have been confirmed around evolved stars, and five of them were detected to orbit giant stars with $\log g < 3.5 $ and $R > 5 R_{\odot}$:
Kepler-91 is detected by transiting, having a hot Jupiter with a semimajor-axis of $0.0731\ \rm{au}$ and orbital period of $6.24658\ \rm{d}$. 
TYC 3667-1280-1 \citep{Niedzielski2016} and 24 Boo \citep{Takarada2018} host warm Jupiters with semimajor-axes of $\sim 0.2\ \rm{au}$ and orbital periods of $\sim 30\ \rm{d}$.
8 UMi \citep{Lee2015} and HIP 67851 \citep{Jones2015} host planets with semimajor-axes of $\sim 0.2\ \rm{au}$ and orbital periods of $\sim 90\ \rm{d}$.
In addition, there are another 13 planets orbiting evolved stars of $\log g < 3.5 $ and $R > 3 R_{\odot}$ with periods shorter than 100 days,
e.g. \textit{Kepler}-56 b and c \citep{Huber2013}, \textit{Kepler}-432 b \citep{Quinn2015}, \textit{K2}-97 b \citep{Grunblatt2016}, \textit{K2}-132 b \citep{Grunblatt2017}, TOI-2269 b \citep{Grunblatt2022}, HD 33142 d \citep{Trifonov2022}.

In addition, based on the latest research on the occurrence rate of planets around evolved stars with radii of $\sim$3--8$R_{\odot}$ given by \citet{Grunblatt2019} and \citet{Pereira2022}, they illustrated that the occurrence of hot Jupiters around these stars is in agreement with the hot Jupiter occurrence rate determined for main sequence stars. 
Thus it can be expected that the lack of planets found in evolved stars with large radii ($R \gtrsim 5 R_{\odot}$) may be due simply to observational bias rather than an actual deficit of planets.

In this paper, we present the discovery of a giant planet orbiting a deeply evolved solar-mass G-type giant star (HD 167768: $\log g = 2.50$) on a close-in orbit ($a = 0.1512\ \rm{au}$) from radial velocity (RV) measurements at Okayama Astrophysical Observatory (OAO). In Section \ref{sec:star} and \ref{sec:observation} we present stellar properties and observations. In Section \ref{sec:orbfit} and \ref{sec:line} we solve the orbit and analyze the line profile and chromospheric activity. Finally, in Section \ref{sec:discuss}, we discuss the results and summarize this work.

% ------------------------------------------------------------------------------------
\section{Stellar properties}\label{sec:star}
\begin{table}
\tbl{Stellar Properties of HD 167768}{
\begin{tabular}{lrr}
\hline\hline
Parameter & Values & Source\\ 
\hline 
$\pi\ (\rm{mas})$& $9.2289$ & \textit{Gaia} EDR3 \\ 
$V$& $6.00$  & \textit{Hipparcos}  \\ 
$B-V$& $0.89$  & \textit{Hipparcos}  \\ 
Spec. type & G8 III & \citet{Houk1999} \\ 
$T_{\mathrm{eff,sp}}\ ({\mathrm{K}})$ & $4830$  & Spectroscopy$^{*}$ \\ 
$\mathrm{[Fe/H]}_{\mathrm{sp}} \mathrm{(dex)}$ & $-0.75$ & Spectroscopy$^{*}$  \\ 
$\log g_{\star,sp}\ ({\mathrm{cgs}})$ & $2.49$  & Spectroscopy$^{*}$ \\ 
$T_{\mathrm{eff}}\ ({\mathrm{K}})$ & $4851_{-44}^{+45}$  & \texttt{isochrones}$^{*}$  \\ 
${\mathrm{[Fe/H]\ (dex)}}$ & $-0.67_{-0.08}^{+0.09}$  & \texttt{isochrones}$^{*}$  \\ 
$\log g_{\star}\ ({\mathrm{cgs}})$ & $2.50_{-0.06}^{+0.06}$  & \texttt{isochrones}$^{*}$  \\ 
$L_{\star}\ (L_{\odot})$ & $46.7_{-2.0}^{+2.0}$  & \texttt{isochrones}$^{*}$  \\ 
$M_{\star}\ (M_{\odot})$ & $1.08_{-0.12}^{+0.14}$  & \texttt{isochrones}$^{*}$  \\ 
$R_{\star}\ (R_{\odot})$ & $9.70_{-0.25}^{+0.25}$  & \texttt{isochrones}$^{*}$  \\ 
Age (Gyr) & $5.31_{-1.68}^{+2.47}$  & \texttt{isochrones}$^{*}$  \\ 
\hline
\end{tabular}}
\begin{tabnote}
\hangindent6pt\noindent
\hbox to6pt{\footnotemark[$*$]\hss}\unskip% 
Determined this work.
\end{tabnote}
\label{tab:star}
\end{table}
\begin{figure}
\begin{center}
\includegraphics[scale=0.5]{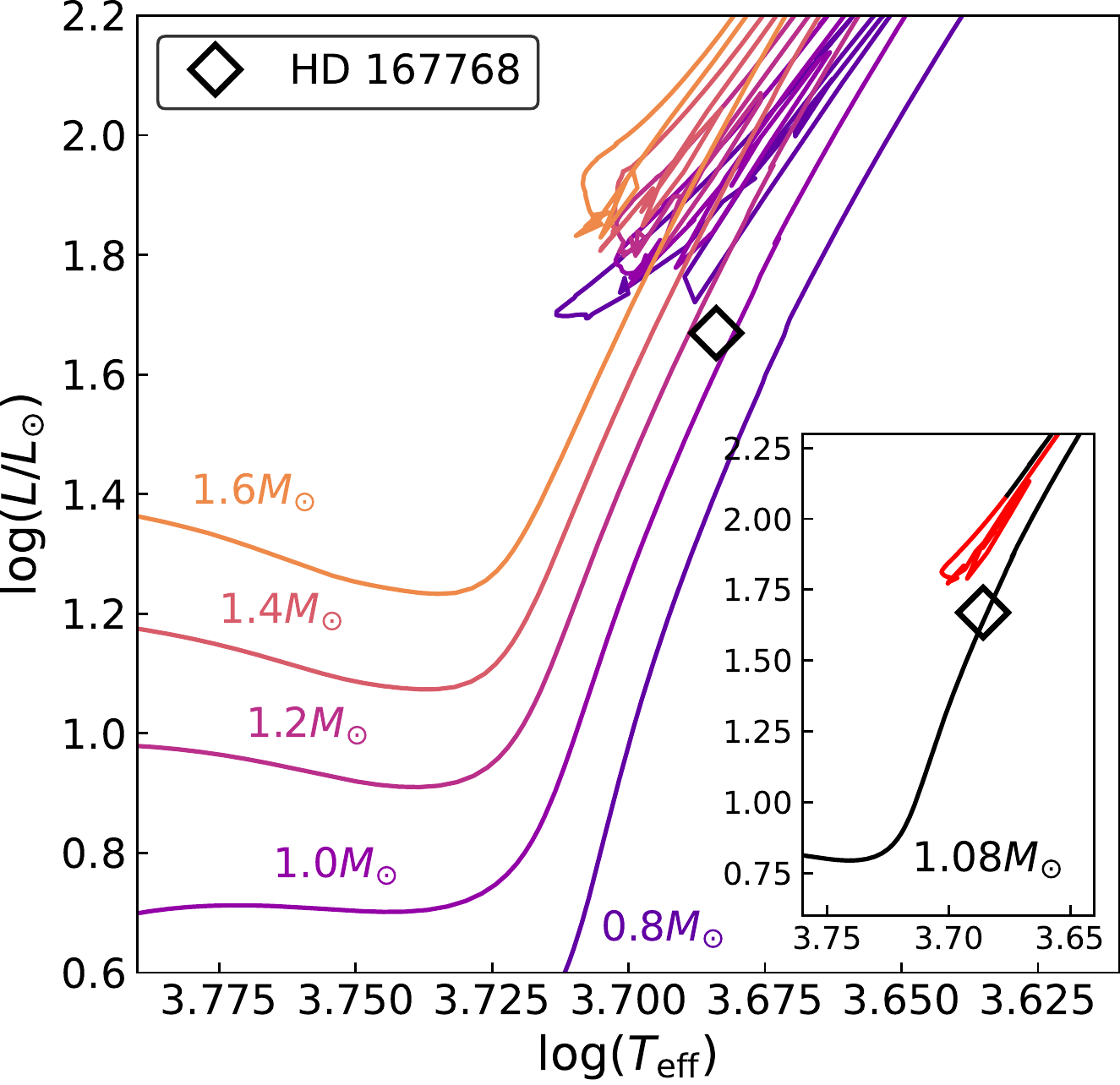} 
\end{center}
\caption{
HR diagram for HD 167768. The star is marked by a open diamond. Evolution tracks of stars having masses between 0.8 and 1.6 $M_{\odot}$ with HD 167768 metallicity ($\mathrm{[Fe/H]} = -0.67$) are shown in solid lines using different colors in the main figure. 
(A colored version of this figure is available in the online journal.)
}\label{fig:HR}
\end{figure}
\begin{figure}
\begin{center}
\includegraphics[scale=0.4]{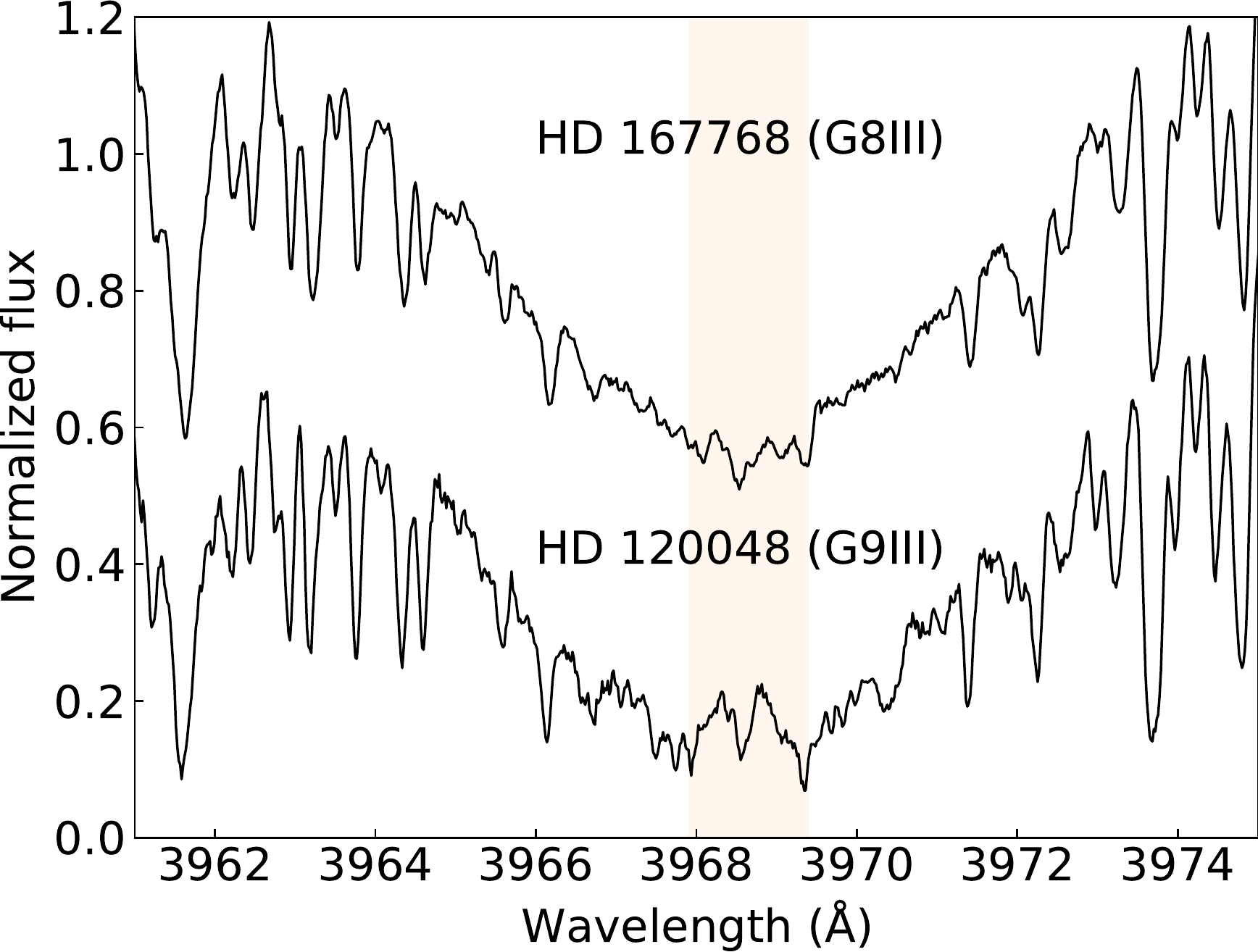}
\end{center}
\caption{
Spectra in the region of Ca \emissiontype{II} H lines of HD 167768 with an active star HD 120048 as a comparison. 
Vertical offsets are added to each normalized spectrum for clarity, and a shaded area is used to mark the core region.
}\label{fig:CaIIH}
\end{figure}
HD 167768 (HR 6840, HIP 89587) is listed in the \textit{Hipparcos} catalog \citep{ESA1997} with apparent magnitude $V$-band $V=6.00$ and classified as G8 III star based on \citet{Houk1999}.
We obtained its parallax of $\pi = 9.2289 \pm 0.0392$ mas from \textit{Gaia} EDR3 \citep{Gaia2016,Gaia2021,Lindegren2021}, and it resulted in a distance of $108.355_{-0.341}^{+0.344}$ pc. 
We determined the atmospheric parameters (effective temperature $T_{\rm{eff,sp}}$, surface gravity $\log g_{\rm{sp}}$, and Fe abundance [Fe/H]$_{\rm{sp}}$) by measuring equivalent width of Fe \emissiontype{I} and Fe \emissiontype{II} lines of iodine-free stellar spectra. 
For star HD 167768, we obtained $T_{\mathrm{eff,sp}}=4830\  {\mathrm{K}}$, $\mathrm{[Fe/H]}_{\mathrm{sp}} = -0.75\ \mathrm{(dex)}$, $\log g_{\star,sp} = 2.49\ ({\mathrm{cgs}})$.
The star is classified as a thick-disk star in \citet{Takeda2008}, showing an enhancement of alpha elements with $\rm{[\alpha/Fe]}=0.26$ (using the average of the silicon and titanium abundance as the alpha-abundance).

We then used \texttt{isochrones} package with grids of \texttt{MIST} stellar evolutionary model to globally fit stellar parameters including stellar luminosity, radius $R_{\star}$, stellar mass $M_{\star}$, and stellar age, as well as atmospheric parameters.  
The best fit were derived from the posterior of a nested sampling with \texttt{PyMultiNest} package.
Consequently, we obtained $L_{\star}= 46.7_{-2.0}^{+2.0} L_{\odot}$, $R_{\star} = 9.70_{-0.25}^{+0.25} R_{\odot}$, and $M_{\star}=1.08_{-0.12}^{+0.14} M_{\odot}$.
Besides, we estimated a maximum rotational period of $P_{\rm{rot}}/\sin i = 111\ \rm{d}$ from the projected rotational velocity of star $v \sin i = 4.44\mathrm{km}\ \mathrm{s}^{-1}$ \citep{Takeda2008} and the star's radius.
A complete listing of stellar properties is given in Table \ref{tab:star}, and a plot in the HR diagram is given in Figure \ref{fig:HR}.

HD 167768 is stable in \textit{Hipparcos} $V$-band photometry with a level of $\sigma_{\rm{HIP}} = 0.007\ \rm{mag}$ (ESA: \cite{vanLeeuwen2007}) among three years, as well as All Sky Automated Survey (ASAS-3) $V$-band photometry with a level of $\sigma_{\rm{ASAS-3}} = 0.03\ \rm{mag}$ \citep{Pojmanski1997} among 7.3 years.
This star is also chromospherically inactive with no significant emission in the core of the Ca \emissiontype{II} H lines. We show its spectrum in Figure \ref{fig:CaIIH} together with another spectrum of a chromospherically active G-type giant star HD 120048. 

% -------------------------------------------------------------------------------------
\section{Observations and RV measurements}\label{sec:observation}
In this work, all the spectra of HD 167768 were obtained from the 1.88-m reflector with HIgh Dispersion Echelle Spectrograph (HIDES: \cite{Izumiura1999}) at Okayama Astrophysical Observatory. 
Its first spectrum was taken in 2004 March under the Okayama Planet Search Program \citep{Sato2005}, an extensive planet survey focusing on RV measurements to late-G (including early-K) giant stars. 
During the 18-year observations, the instrument had been upgraded several times.
In December 2007, the CCD of HIDES was upgraded from the single one to a mosaic of three, which widened the wavelength region from 5000-6100 \AA ~to 3700-7500 \AA, and enabled us to simultaneously measure the level of stellar activities (e.g. Ca \emissiontype{II} H lines) and line profiles as well as radial velocities.
In 2010, a new high-efficiency fiber-link system with its own iodine cell was installed on the HIDES, which greatly enhanced the overall throughput \citep{Kambe2013}. 
In 2018, another upgrade was carried out to enhance the performance of the fiber-link system with a newly designed optical path and stabilizing platform\footnote{A more detailed introduction to the HIDES fiber-link mode upgrade and its performance will be given in a forthcoming paper.}. In this upgrade, the slit mode elements were entirely removed from the HIDES optical path.
Hereafter, we name the observation taken by conventional slit mode, fiber mode pre-upgrade in 2018, and fiber-mode post-upgrade in 2018 as HIDES-S, HIDES-F1, and HIDES-F2, respectively.

In the case of HIDES-S observations, the slit width was set to 200 $\mu$m ($\timeform{0.76''}$) corresponding to the resolution $\mathit{R} = \lambda / \Delta \lambda \sim 67000$ by about 3.3-pixel sampling. The typical exposure time and signal-to-noise ratio (S/N) for HIDES-S is 1200 seconds and over 150, respectively.
In the case of HIDES-F1 and -F2 observations, the width of the sliced image was $\timeform{1.05''}$ corresponding to the resolution $\mathit{R} \sim 55000$ by about 3.8-pixel sampling. The typical exposure time and S/N for both HIDES-F1 and -F2 are 600 seconds and over 200, respectively.
We adopted the data which gained S/N over approximately 100 per pixel at $\sim$5500 \AA ~within 1800 seconds. Finally, we collected 38, 36, and 28 spectra from HIDES-S, F1, and F2, respectively, between 2004 March and 2022 May.

The reduction of the echelle spectra was performed with \texttt{IRAF} in a standard manner. 
However, the 3700--4000 \AA ~region, including Ca \emissiontype{II} H lines region, of HIDES-F1 and -F2 spectra were deprecated due to the aperture overlaps.
For precise RV measurements, we used spectra covering 5000 to 5800 \AA ~in which I$_{\rm{2}}$ absorption lines are superimposed by an I$_{\rm{2}}$ cell. We computed RV variations following the method described in \citet{Sato2002} and \citet{Sato2012} which was based on a method by \citet{Butler1996a}. 
We modeled the spectra (I$_{\rm{2}}$ superposed stellar spectra) by using the stellar template. The stellar template spectra used for HIDES-S were obtained by deconvolving pure stellar spectra with the IP estimated from I$_{\rm{2}}$ superposed B type star or flat spectra. The stellar template spectra used for HIDES-F1 and -F2 were obtained by HIDES-F1 High-resolution mode ($\mathit{R} \sim 100000$) observations without I$_{\rm{2}}$ in its optical path. Instrumental RV offsets were assumed between HIDES-S, -F1, and -F2.

% -------------------------------------------------------------------------------------
\section{Orbit fitting and planetary parameters}\label{sec:orbfit}
\begin{figure}
\begin{center}
\includegraphics[scale=0.6]{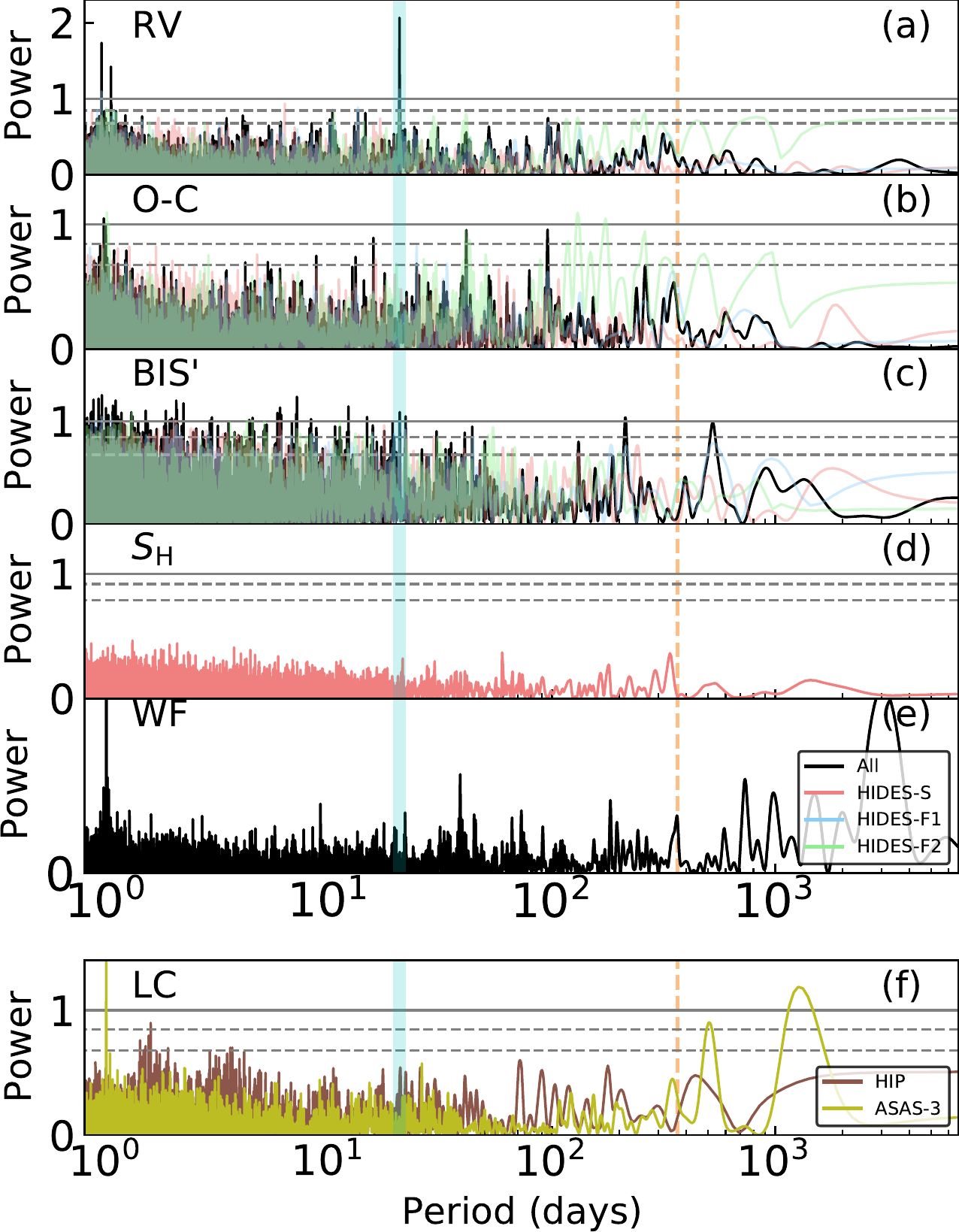} 
\end{center}
\caption{
Generalized Lomb-Scargle (GLS) periodograms for star HD 167768. 
In all subplots, vertical axes are given in normalized power \citep{Baluev2008} corresponding to 0.1\% FAP level, and horizontal axes are given in period (days). 
The subplots respectively represent (a) the GLS periodograms of the observed RVs, (b) the residuals to single Keplerian orbital fit, (c) BIS', (d) $S_{\rm{H}}$, (e) window function, and (e) light curves,
The horizontal grey lines represent 10\%, 1\%, and 0.1\% FAP levels from bottom to top, and particularly the 0.1\% FAP level is marked in solid line. 
The vertical cyan solid lines indicate the best-fitted period from the single Keplerian model, and the vertical orange dashed line indicates one year. 
 (A colored version of this figure is available in the online journal.)}\label{fig:ls}
\end{figure}
\begin{figure}
\begin{center}
\includegraphics[scale=0.55]{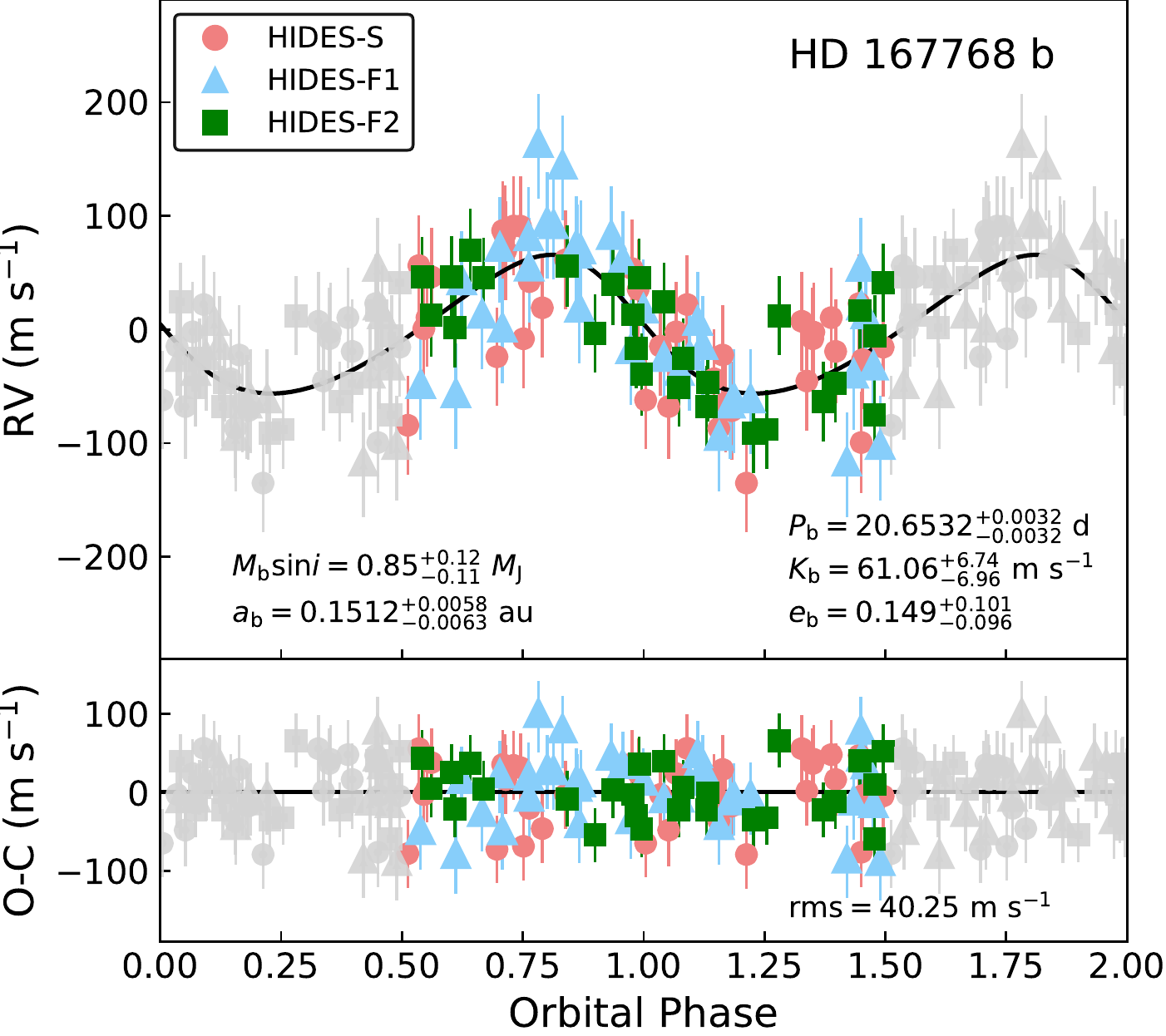} 
\end{center}
\caption{
Phase folded best-fit orbital solution of HD 167768. 
Upper panel: The best-fit orbital solution from the MCMC fitting.
The black solid line shows the best-fit single Keplerian curve. The colored dots with errorbars are observed RVs in one orbital phase with fitted RV offsets shifted between instruments and with fitted jitters quadratically added to the observational errorbars. The non-colored dots are in extended half phases.
Lower panel: The residuals to the best fit in the upper panel. 
In both two subplots, the symbols are the same. HIDES-S data are shown in light red circles, HIDES-F1 data are shown in light blue triangles, and HIDES-F2 data are shown in green squares.
(A colored version of this figure is available in the online journal, and a complete RV data listing will be available online as supplementary after the publication.)}\label{fig:rv}
\end{figure}
\begin{table*}
\tbl{Orbital parameters}{
\begin{tabular}{lcclcc}
\hline\hline
Parameters  & Best-fit & Prior  & Parameters  & Best-fit & Prior\\ 
\hline 
$P\ (\rm{d})$  & $20.6532_{-0.0035}^{+0.0034}$ & Jeffery's $(1, 5000)$ & $s_{\rm{s}}\ (\rm{m\ s^{-1}})$ & $42.70_{-5.37}^{+5.71}$ & Mod-Jeffery's $(1.01 (1), 100)$ \\ 
$K\ (\rm{m\ s^{-1}})$ & $61.06_{-6.96}^{+6.74}$ & Mod-Jeffery's $(1.01 (1), 1000)$  & $s_{\rm{f1}}\ (\rm{m\ s^{-1}})$ & $45.52_{-5.08}^{+6.12}$ &  Mod-Jeffery's $(1.01 (1), 100)$ \\ 
$\sqrt{e}\cos \omega$& $0.1849_{-0.3039}^{+0.1841}$ & Uniform $(-1, 1)$ & $s_{\rm{f2}}\ (\rm{m\ s^{-1}})$ & $33.30_{-5.19}^{+6.16}$ &  Mod-Jeffery's $(1.01 (1), 100)$\\ 
$\sqrt{e}\sin \omega$ & $0.3636_{-0.3074}^{+0.0468}$ & Uniform $(-1, 1)$ & $\gamma_{\rm{s}}\ (\rm{m\ s^{-1}})$ & $1.59_{-7.55}^{+7.73}$ & Uniform $(-500, 500)$ \\  
$T_{\rm{c}}$ (BJD$-2450000$) & $4595.44_{-0.73}^{+0.67}$ & Uniform $(3078, 9731)$ & $\gamma_{\rm{f1}}\ (\rm{m\ s^{-1}})$ & $-9.55_{-7.78}^{+7.91}$ & Uniform $(-500, 500)$ \\ 
$e$ & $0.149_{-0.096}^{+0.101}$ & (derived) & $\gamma_{\rm{f2}}\ (\rm{m\ s^{-1}})$ & $7.15_{-7.40}^{+7.09}$ & Uniform $(-500, 500)$ \\ 
$\omega\ \rm{(rad)}$ & $1.025_{-0.790}^{+0.865}$ & (derived) & --------- & --------- & --------- \\ 
$T_{\rm{p}}$ (BJD$-2450000$) & $4594.07_{-2.25}^{+3.10}$ & (derived) & $\rm{rms}\ (\rm{m\ s^{-1}})$ & $40.25$ &  (derived) \\ 
$M_{\rm{p}}\sin i\ (M_{\rm{J}})$ & $0.85_{-0.11}^{+0.12}$ & (derived) & $\chi^{2}_{\rm{red}}$ & $1.010$ &  (derived) \\ 
$a\ (\rm{au})$ & $0.1512_{-0.0063}^{+0.0058}$ & (derived) & $\rm{BIC}$ & $1091.7$ & (derived) \\ 
\hline
\hline
\end{tabular}}
\begin{tabnote}
The subscript ``s'', ``f1'', and ``f2'' refer to HIDES-S, -F1, and -F2 data respectively. Modified Jeffery's prior has $(\min\ (\rm{knee\ value}),\ \max)$
\end{tabnote}
\label{tab:orb}
\end{table*}

First, we performed a Generalized Lomb-Scargle periodogram (hereafter GLS: \cite{Zechmeister2009}, Figure \ref{fig:ls}) to search for periodicity in the time series and calculated False Alarm Probability (FAP) to assess the significance of the periodicity.
We considered a peak in GLS with FAP significantly lower than 0.1\% as a credible signal for regular variation.

The Keplerian orbital fit was finished similarly to our previous works \citep{Teng2022}.
We generated Keplerian model and priors with $\mathtt{RadVel}$ package. 
The fitted Keplerian orbital elements includes orbital period $P$, RV semi-amplitude $K$, the combination of eccentricity $e$ and argument of periastron $\omega$, $\sqrt{e} \cos \omega$ and $\sqrt{e} \sin \omega$, and time of inferior conjunction. The extra jitter $s$ and RV offset to zero point $\gamma$ were also set as free parameters, and the pariastron passage $T_{\rm{p}}$ was derived from inferior conjunction $T_{\rm{c}}$.

The best-fit Keplerian orbit and its uncertainties were derived by the maximum a posteriori (MAP) method from Markov Chain Monte Carlo (MCMC) sampling with $\mathtt{emcee}$ package. 
Model selection adopted reduced Chi-square $\chi^{2}_{\rm{red}}$ and Bayesian Information Criteria (BIC: \cite{Schwarz1978}).  

For HD 167768 RVs, we could find a strong signal at 20 d, indicating a regular variation in the time series.
We then fitted a single Keplerian to the data in the way above.
Consequently, we obtained orbital parameters of $P = 20.6532_{-0.0032}^{+0.0032}\ \rm{d}$, $K = 61.06_{-6.96}^{+6.74}\ \rm{m\ s^{-1}}$, and $e = 0.149_{-0.097}^{+0.092}$. Adopting a stellar mass of $1.08 M_{\odot}$, we obtained a minimum mass of $M_{\rm{p}}\sin i = 0.85_{-0.11}^{+0.12}\ M_{\rm{J}}$ and a semimajor axis of $a = 0.1512_{-0.0063}^{+0.0058}\ \rm{au}$ for the planet. 
The best-fit single Keplerian model has a decreased BIC value of 60 compared to the No-planet model, suggesting the significance of the RV variation. 
A detailed parameter listing is given in Table \ref{tab:orb}, and the phase-folded orbit is shown in Figure \ref{fig:rv}.

\begin{figure}
\begin{center}
\includegraphics[scale=0.45]{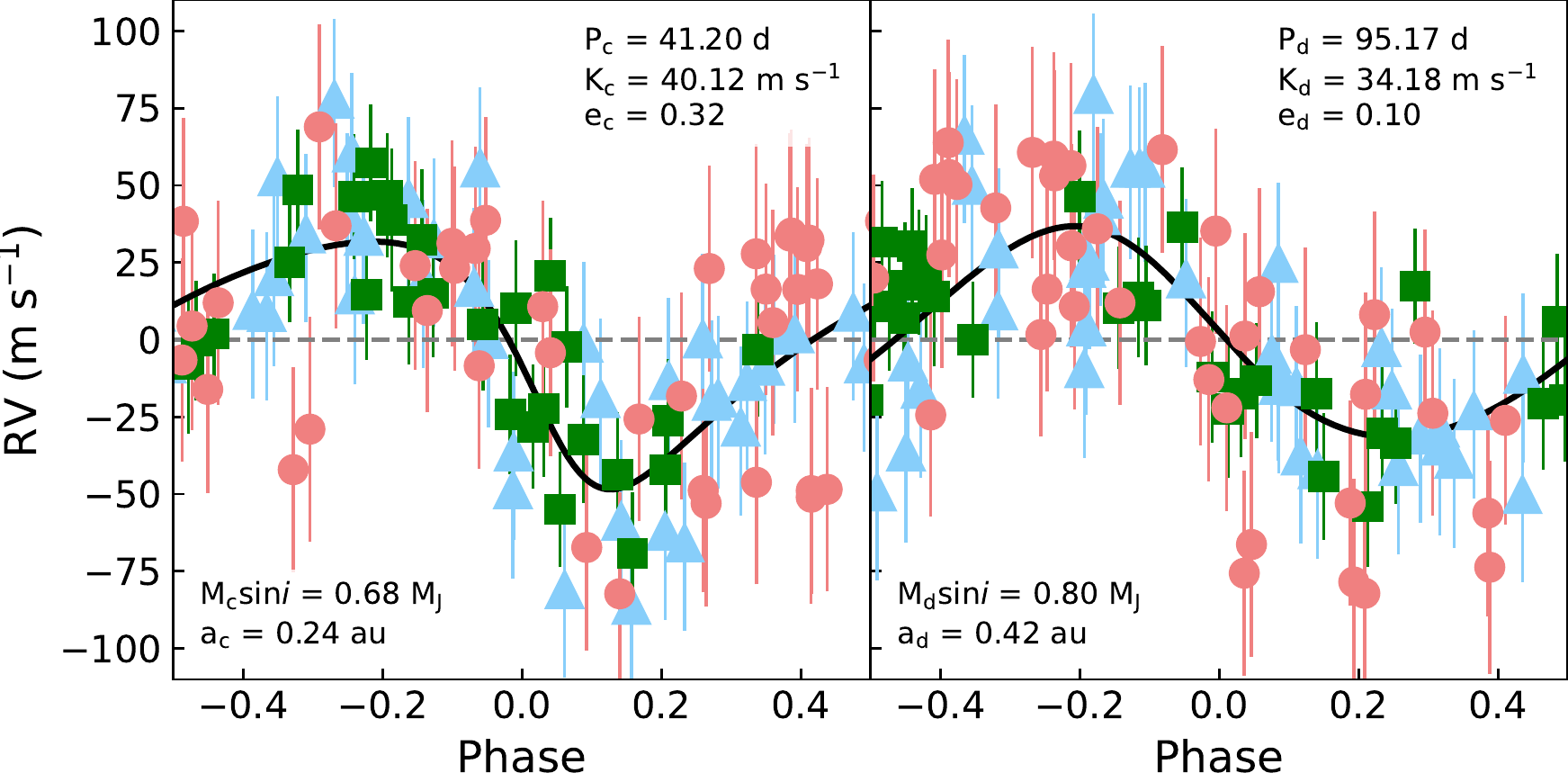} 
\end{center}
\caption{
Phase folded best-fit orbital solution of the possible extra companions in HD 167768 system. The symbols are the same as those in Figure \ref{fig:rv}.
(A colored version of this figure is available in the online journal, and a complete RV data listing will be available online as supplementary after the publication.)}\label{fig:rv3}
\end{figure}

The residuals to single Keplerian fit has rms of $40.25\ \rm{m\ s^{-1}}$, which is significant larger than the estimated $p$-mode oscillation jitter of $10.1\ \rm{m\ s^{-1}}$ \citep{Kjeldsen1995} or $20.4\ \rm{m\ s^{-1}}$ \citep{Kjeldsen2011}. 
In the periodogram of the residuals, there are two additional signals at 41 d and 95 d with FAP slightly lower than 0.1\%, suggesting possible extra companions in the system. 
Keplerian fits yielded semi-amplitudes of $\sim 40\ \rm{m\ s^{-1}}$ and $\sim 34\ \rm{m\ s^{-1}}$ for the periodicity, corresponding to two planets of $0.68 M_{\rm{J}}$ at $0.24\ \rm{au}$ and $0.80 M_{\rm{J}}$ at $0.42\ \rm{au}$, respectively (Figure \ref{fig:rv3}). 
The rms of three-Keplerian is reduced to $\sim\ 26\ \rm{m\ s^{-1}}$, and BIC values is reduced to $1044$, but the goodness-of-fit of $\chi_{\rm{red}}^{2}=1.056$ has larger deviation to unity compared to the single Keplerian result. 
Thus, continuous monitoring will enable us to validate the periodicity.

% -------------------------------------------------------------------------------------
\section{Line profile and chromospheric activity}\label{sec:line}
\begin{figure}
\begin{center}
\includegraphics[scale=0.55]{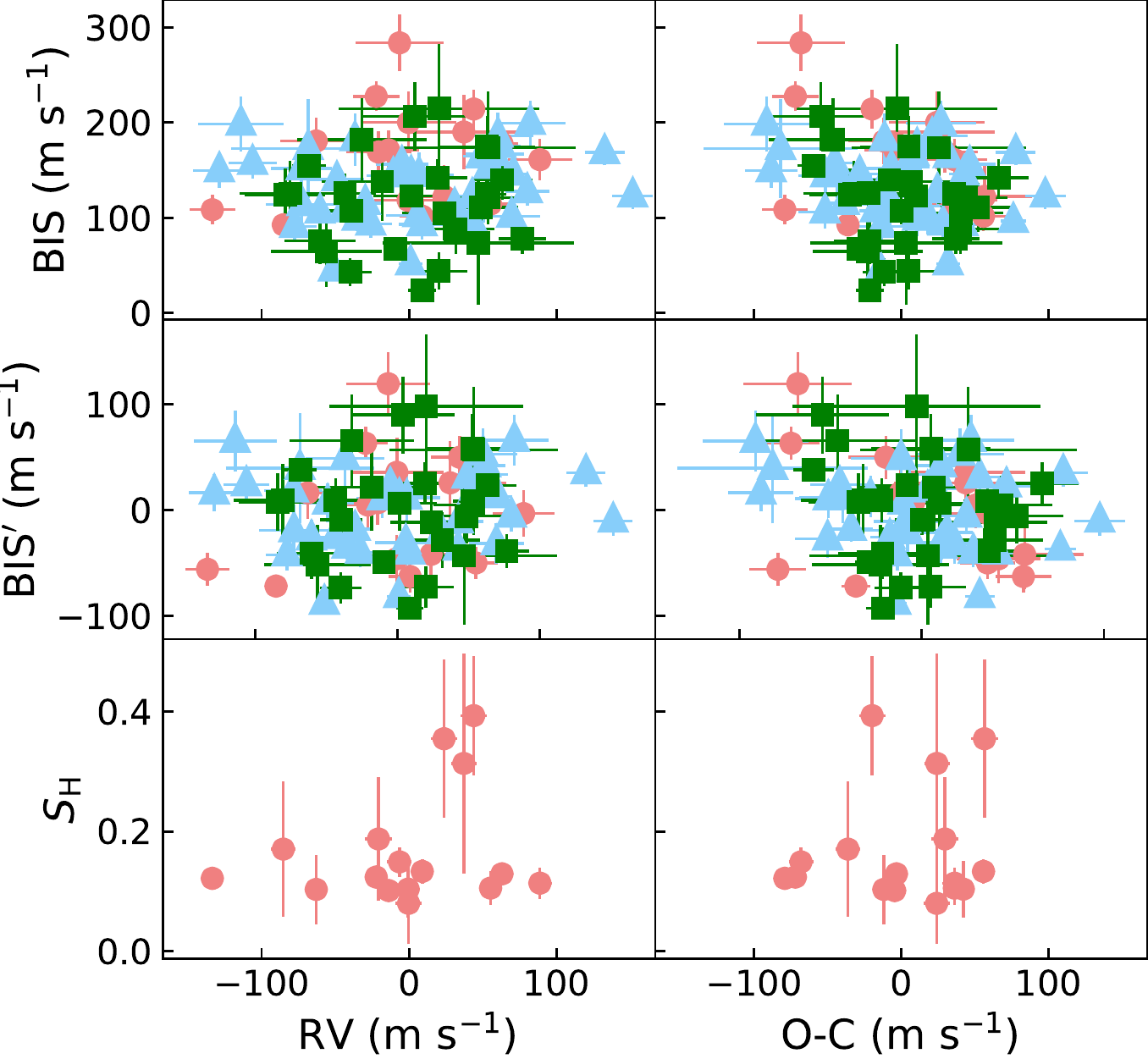} 
\end{center}
\caption{
BIS, mean removed BIS (BIS'), and index of Ca \emissiontype{II} H line ($S_{\rm{H}}$) respectively against the observed RVs and the residuals to best-fit orbit (O-C).
In all subplots, the symbols are the same as those in Figure \ref{fig:rv}. 
(A colored version of this figure is available in the online journal.)
}\label{fig:line}
\end{figure}

We also calculated the deformation spectral line profiles and the flux of the chromospheric Ca \emissiontype{II} H line cores because line profile deformation and stellar chromospheric activity can both result in RV variations and mimic planetary signals \citep{Queloz2001}.

We follow the methodology in \citet{Takarada2018} and \citet{Teng2022}.
We use Bisector inverse span (BIS: \cite{Dall2006}) as an indicator of line profile asymmetry. 
The BIS was defined as the velocity offset between the upper region and lower region in the cross-correlation function (CCF: \cite{Baranne1996,Pepe2002}).
Since different instruments show different instrumental profiles, we define the mean removed BIS ($\rm{BIS}^{\prime}=\rm{BIS}-\overline{BIS}$) to suppress the difference between instruments. 
The BIS calculations were based on the iodine-free spectra in the range of 4000–5000 \AA. 
We also applied the Ca \emissiontype{II} H index ($S_{\rm{H}}$: defined in \cite{Sato2013}) to quantify the strength of chromospheric activity \citep{Duncan1991}. 
A clear correlation between RV and $S_{\rm{H}}$ can be detected (e.g., HD 120048 in Figure 9 in \cite{Sato2013}) for a chromospheric active star having RV variate with $S_{\rm{H}}$ simultaneously. 

Illustrated in Figure \ref{fig:line}, the BIS of HIDES-S, -F1, and -F2 have almost the same level of line dispersion, suggesting that stellar surface modulation should greatly affect the spectral line profile. 
However, BIS$^{\prime}$ has no correlation with RV ($r=0.12$). Similarly, $S_{\rm{H}}$ also has weak correlation with RV ($r=0.24$). Furthermore, we calculated periodogram of BIS$^{\prime}$ and $S_{\rm{H}}$ (Figure \ref{fig:ls}). 
Concerning the bundle of powerful high-frequency signals, the 20.7 d signal that appears in BIS$^{\prime}$ periodogram is not considered to be significant. 
The periodogram of $S_{\rm{H}}$ as well as HIP and ASAS-3 light curves do not show any periodicity around the 20.7 d signal. Thus we conclude orbital motion is the source of 20.7 d RV variation. 

\section{Discussion and summary}\label{sec:discuss}
We have reported the discovery of a giant planet orbiting the G type giant star HD 167768 ($1.08 M_{\odot}$, $9.70 R_{\odot}$, and $\log g = 2.50$) on a close-in orbit ($P = 20.6532\ \rm{d}$, $a = 0.1512\ \rm{au}$) from radial velocity (RV) measurement at OAO. The planet, HD 167768 b, is known to have the shortest period and proximity to a central star having $R \gtsim 10 R_{\odot}$.

Given the planet's orbital parameters and star's parameters, we can derive a transit probability of 35\% and a maximum transit duration of 2.3 d (55 hr).
The star is luminous and it results in a high planet's equilibrium temperature of $T_{\rm{eq}} = 1874\ \rm{K}$ with assumed Bond albedo $A$ equal to 0:
\begin{equation}
    T_{\rm{eq}} = T_{\rm{eff}}(1-A)^{1/4}\sqrt{\frac{R_{\star}}{2a}}.
\end{equation}
The planet is the hottest Jupiter found by RV measurements, and it is comparable to the hottest giant planets ever detected\footnote{NASA Exoplanet Archive \url{https://exoplanetarchive.ipac.caltech.edu} \citep{Akeson2013}}.
High irradiation for the expanding giant star could lead to an inflation of the warm Jupiter like HD 167768 \citep{Lopez2016}. 
Theoretically, a gas giant planet of $0.85 M_{\rm{J}}$ orbiting a giant star of $1 M_{\odot}$ and $10 R_{\odot}$ with orbital period of $20\ \rm{d}$ could be inflated to approximately $2 R_{\rm{J}}$ with interior heating efficiency of 0.5\% (Figure 8 in \cite{Lopez2016}).
However, based on the actual measurement of the planet re-inflation efficiency (\textit{K2}-97 b and \textit{K2}-132 b: \cite{Grunblatt2016, Grunblatt2017}), it is one order of magnitude lower than the theoretical prediction. 
Simply, if we assume a planet radius of $1.5 R_{\rm{J}}$, we can expect a transit depth of about 0.02\%, which is within the detectability of space telescopes. 
But unfortunately, this star is not included in the \textit{TESS} observation plan by 2023 September 20th\footnote{\url{https://heasarc.gsfc.nasa.gov/cgi-bin/tess/webtess/wtv.py}}.

HD 167768 system is extremely compact concerning that the orbital separation at pericenter is only $a(1-e) = 0.128\ \rm{au}$, that is $2.6 R_{\star}$, and it is expected to have experienced tidal orbital decay \citep{Villaver2014}. 
The average $(a/R_{\star}) \sim 3$ is also one of the smallest among the known planetary systems \citep{Patra2020}.
Here, by assuming the reduced tidal quality factor $Q_{\star}^{\prime} \sim 10^6$ and equilibrium tide formulations in \citet{Patra2020}, we can estimate that tidal inspiral timescale of HD 167768 b is approximately 0.5 Gyr.

The moderate eccentricity of HD 167768 b ($e = 0.149$) is indeed in line with the trend of detected close-in giant planets around evolved stars. \citet{Grunblatt2018} suggested that this trend results from the longer tidal circularization timescale than the tidal orbital decay timescale and thus the planets with moderate eccentricities are transient objects. However, we cannot estimate the orbit circularization timescale for two reasons. One is that the radius of the planet is unknown and \citep{Patra2017}, another one is that we are tentatively not sure if the system consists of outer companions (Section \ref{sec:orbfit}).
Detailed analysis of the eccentricity evolution in future work is highly encouraged.

\begin{figure}
\begin{center}
\includegraphics[scale=0.55]{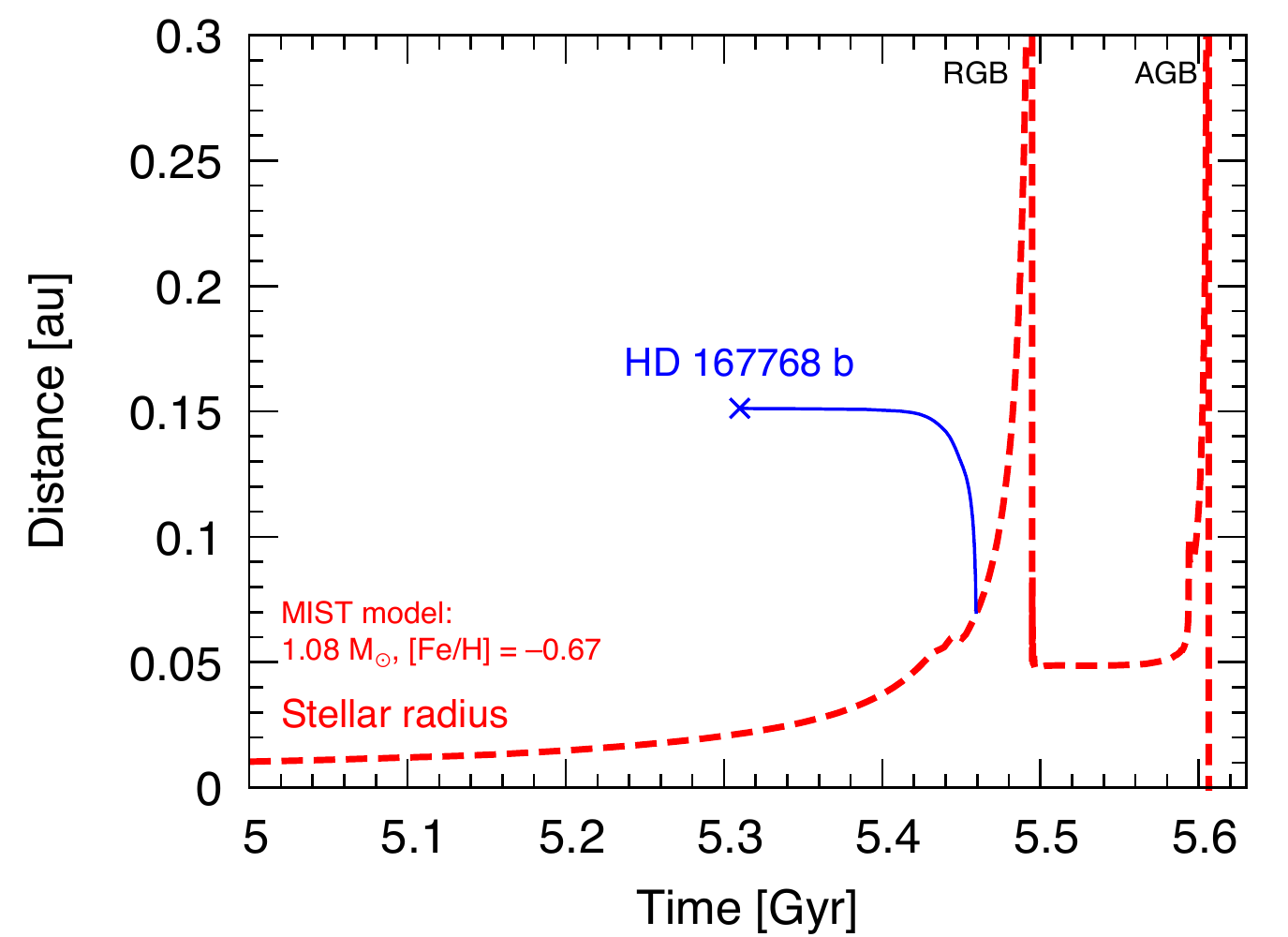} 
\end{center}
\caption{
Evolutions of the planet's semimajor axis (solid) and stellar radius (dashed).
(A colored version of this figure is available in the online journal.)
}\label{fig:orbevol}
\end{figure}

As the star is expected to be ascending the red giant branch (RGB; see Figure \ref{fig:HR}), the ultimate fate of the planet should be an engulfment. 
\citet{Sun2018} predicted the critical radius $a_{\rm{crit}}(t)$ of main stars in binarities, where $a_{\rm{crit}}(t)$ is defined as the initial separation out to which orbits would have decayed down to the surface of the primary by the age $t$.
For a main star of $1M_{\odot}$, all the observed systems have $a>a_{\rm{crit}}(t)$, indicating the orbital decay rate of the observed systems is small compared to the stellar evolution time-scale.
HD 167768 system has $a=0.1512\ \rm{au}$ and $\log g = 2.5$, which also well obeys the trend shown by the Figure 16 in \citep{Sun2018}. 
Nevertheless, as HD 167768 will quickly expand in RGB, its planet will be swallowed in a short time.
We simulated the orbital evolution of the planet following \citet{Kunitomo2011} and using the \texttt{MIST} stellar evolution model\footnote{Since the internal structure information is not available in the \texttt{MIST} model, we assumed that the star has a fully convective structure. We confirmed that this assumption does not have a significant impact on the result of orbital evolution.} as
\begin{equation}
    \frac{1}{a} \frac{d a}{d t}=-6 \frac{k}{T} \frac{M_{\rm p}}{M_{\star}}\left(1+\frac{M_{\rm p}}{M_{\star}}\right)\left(\frac{R_{\star}}{a}\right)^8-\frac{\dot{M}_{\star}}{M_{\star}}\,.
\end{equation}
We refer the readers to Equations (2)--(5) of \citet{Kunitomo2011} for the tidal model with the $k$ and $T$ parameters and the stellar mass-loss rate $\dot{M}_{\star}$.
Figure \ref{fig:orbevol} shows that the planet will be engulfed by the expanding star within 0.15\,Gyr.

Moreover, HD 167768 system and the 24 Boo system (\cite{Takarada2018}; Teng et al. in prep) are almost twins with several similarities:

In terms of the host stars, they are solar-mass giant stars at similar evolution stages and both reside in the thick disk \citep{Takeda2008} in the galaxy. 
\citet{Bashi2022} have found that, compared to thin-disc stars, planets are less preferable around thick-disc stars, which are commonly known as more iron-poor and alpha enhanced stars. Recently, several typical thick-disk stars were confirmed to be planet-hosts by \textit{TESS}, e.g. LHS 1518 ($\rm{[Fe/H]} = -0.12$: \cite{Gan2020}) and TOI-561 ($\rm{[Fe/H]} = -0.41$: \cite{Weiss2021}).

They are both iron-poor with $\rm{[Fe/H]} \lesssim -0.70$ but with enhancement of alpha elements \citep{Takeda2008}, being in agreement with the tendency that iron-poor planet hosts are generally enhanced more in alpha elements \citep{Haywood2009,Adibekyan2012,Bashi2022}. 
In the contrast, a number of planets were found around evolved stars that had particularly high metallicities (e.g. recent discoveries by \citet{Moutou2021,Grunblatt2022}). These close-in planets stand on the two opposite side of the well-known planet-metallicity relations \citep{Fischer2005,Petigura2018}.

Regarding the planetary systems, they both host warm Jupiters in close-in orbits with $a \lesssim 0.2\ \rm{au}$, and they both have extra RV variations in their residuals to the best-fit orbital solution, suggesting the existence of possible outer planets.

Are they related to a certain population? Future monitoring and characterization might uncover the truth.

%%%%%%%%%%%%%%%%%%%%%%%%%%%%%%%%%%%%%%%
\begin{ack}
We thank the anonymous referee for the very valuable and important suggestions that allowed us to improve the manuscript.

This research is based on data collected at the Okayama Astrophysical Observatory (OAO), which was operated by the National Astronomical Observatory of Japan. 
We are grateful to all the staff members of OAO for their support during the observations.
The Okayama 188cm telescope is operated by a consortium led by Exoplanet Observation Research Center, Tokyo Institute of Technology (Tokyo Tech), under the framework of tripartite cooperation among Asakuchi-city, NAOJ, and Tokyo Tech from 2018.
We thank the students of Tokyo Institute of Technology and Kobe University for their kind help with the observations at OAO. 
We express our special thanks to Yoichi Takeda for the support in stellar property analysis. 
B.S. was partially supported by MEXT's program ``Promotion of Environmental Improvement for Independence of Young Researchers" under the Special Coordination Funds for Promoting Science and Technology, and by Grant-in-Aid for Young Scientists (B) 17740106 and 20740101, Grant-in-Aid for Scientific Research (C) 23540263, Grant-in-Aid for Scientific Research on Innovative Areas 18H05442 from the Japan Society for the Promotion of Science (JSPS), and by Satellite Research in 2017-2020 from Astrobiology Center, NINS.
H.I. was supported by JSPS KAKENHI Grant Numbers JP16H02169, JP23244038.
Y.J.L is supported by the National Key R\&D Program of China No. 2019YFA0405102.

This research has made use of \textsc{IRAF}. It is distributed by the National Optical Astronomy Observatories, which is operated by the Association of Universities for Research in Astronomy, Inc. under a cooperative agreement with the National Science Foundation, USA.
This research has adopted \texttt{MIST} version 1.2 \citep{Paxton2011, Paxton2013, Paxton2015, Choi2016, Dotter2016, Paxton2018}.
This research has made use of the following \texttt{Python} packages for scientific calculation: \texttt{NumPy} \citep{Harris2020}, \texttt{SciPy} \citep{Virtanen2020} and \texttt{astropy} \citep{Astropy2013}, \texttt{isochrones} \citep{Morton2015}, \texttt{PyMultiNest} \citep{Feroz2008,Feroz2009,Feroz2019}, \texttt{radvel} \citep{Fulton2017, Fulton2018}, and \texttt{emcee} \citep{Foreman-Mackey2013}.

This research has made use of the SIMBAD database, operated at CDS, Strasbourg, France.
This research has made use of data from the European Space Agency (ESA) mission
{\it Gaia} (\url{https://www.cosmos.esa.int/gaia}), processed by the {\it Gaia}
Data Processing and Analysis Consortium (DPAC,
\url{https://www.cosmos.esa.int/web/gaia/dpac/consortium}). Funding for the DPAC
has been provided by national institutions, in particular the institutions
participating in the {\it Gaia} Multilateral Agreement.
This research has made use of the NASA Exoplanet Archive,
which is operated by the California Institute of Technology,
under contract with the National Aeronautics and Space Administration under the Exoplanet Exploration Program.
\end{ack}

%%%
% See the manual for the detail.
%%%

\end{document}